\documentclass[journal]{IEEEtran}
\usepackage{amsfonts}
\usepackage{amsmath}
\usepackage{amssymb}
\usepackage{subfigure}
\usepackage{epsfig}
\usepackage{color}
\usepackage{float}

\usepackage{cite,graphicx,amsmath,amssymb,cite,algorithm}

\begin{document}
\newtheorem{theorem}{Theorem}
\newtheorem{proposition}{Proposition}
\newtheorem{definition}{Definition}
\newtheorem{lemma}{Lemma}
\newtheorem{corollary}{Corollary}
\newtheorem{remark}{Remark}
%\newtheorem{theorem}{Theorem}
%\newtheorem{proposition}{Proposition}
%\newtheorem{definition}{Definition}
%\newtheorem{lemma}{Lemma}
%\newtheorem{corollary}{Corollary}
%\newtheorem{remark}{Remark}
%\newtheorem{construction}{Construction}
%\newtheorem{algorithm}{Algorithm}
%\newcommand{\supp}{\mathop{\rm supp}}
%\newcommand{\sinc}{\mathop{\rm sinc}}
%\newcommand{\spann}{\mathop{\rm span}}
%\newcommand{\essinf}{\mathop{\rm ess\,inf}}
%\newcommand{\esssup}{\mathop{\rm ess\,sup}}
%\newcommand{\Lip}{\rm Lip}
%\newcommand{\sign}{\mathop{\rm sign}}
%\newcommand{\osc}{\mathop{\rm osc}}
%\newcommand{\R}{{\mathbb{R}}}
%\newcommand{\Z}{{\mathbb{Z}}}
%\newcommand{\C}{{\mathbb{C}}}
%=========================================

\title{Sum-MSE performance gain of DFT-based channel estimator over frequency-domain LS one in full-duplex OFDM systems with colored interference}
%====================================================Pilot Optimization,
\author{{Jin Wang, Feng Shu, Jinhui Lu, Hai Yu, Riqing Chen, Jun Li, and Dushantha Nalin K. Jayakody}
\thanks{This work was supported in part by the National Natural Science Foundation of China (Nos. 61271230, 61472190, 6147210, and 61501238), the Open Research Fund of National Key Laboratory of Electromagnetic Environment, China Research Institute of Radiowave Propagation (No. 201500013),and the open research fund of National Mobile Communications Research Laboratory, Southeast University, China (No. 2013D02),}
\thanks{Jin Wang, Feng Shu, Jinhui Lu, Hai Yu, and Jun Li  are with School of Electronic and Optical Engineering, Nanjing University of Science and Technology, 210094, CHINA. E-mail:\{jin.wang, shufeng, juhuilu, haiyu, jun.li\}@njust.edu.cn}
\thanks{Feng Shu and Riqing Chen are with the College of Computer and Information Sciences, Fujian Agriculture and Forestry University, Fuzhou 350002, China. E-mail: riqing.chen@fafu.edu.cn.}
\thanks{Feng Shu is also with the National Key Laboratory of Electromagnetic Environment, China Research Institute of Radiowave Propagation, Qingdao 266107, China.}

%\thanks{Feng Shu is with the College of Computer and Information Sciences, Fujian Agriculture and Forestry University, Fuzhou 350002, China. E-mail: riqing.chen@fafu.edu.cn.}
\thanks{Dushantha Nalin K. Jayakody is with  the Department of Control System Optimization, Institute of Cybernetics, National Research Tomsk Polytechnic University, Russia.}
}
\maketitle
\begin{abstract}
In this paper, we make an investigation on the sum-mean-square-error (sum-MSE) performance gain achieved by DFT-based least-square (LS) channel estimator over frequency-domain LS one in full-duplex OFDM system in the presence of colored interference and noise. The closed-form expression of the sum-MSE performance gain is given. Its simple upper and lower bounds are derived by using inequalities of matrix eigen-values. By simulation and analysis, the upper lower bound is shown to be close to the exact value of MSE gain as the ratio of the number $N$ of total subcarriers to the cyclic prefix length $L$ grows and the correlation factor of colored interference increases. More importantly, we also find that the MSE gain varies from one to $N/L$ as the correlation among  colored interferences decreases gradually. According to theoretical analysis, we also find the MSE gain has very simple forms in two extreme scenarios. In the first extreme case that the colored interferences over all subchannels are fully correlated, i.e., their covariance matrix is a matrix of all-ones, the sum-MSE gain reduces to 1. In other words, there is no performance gain. In the second extreme case that the colored-interference covariance matrix  is an identity matrix, i.e, they are mutually independent, the achievable sum-MSE performance gain is $N/L$. A large ratio $N/L$  will achieve a significant sum-MSE gain.
\end{abstract}

%========================Section I===================================
\section{Introduction}
Recently, full-duplex (FD) technique becomes a hot research field in internet of things (IoT), and wireless networks due to its ability of doubling data transmission rate by  simultaneously transmitting and receiving signals over the same frequency band and time slot compared to time-division-duplex (TDD) and frequency-division-duplex (FDD) mode\cite{Zhang,Riihonen,Xie,Sun,Kim,Riihonen2}. The major  problem of facing FD  is that the weak fading received signal is severely interfered with the strong FD self-interference (SI)\cite{Kim,Riihonen2,Ju}. In \cite{Sabharwal}, the SI cancellation schemes are divided into three categories: propagation-domain, analog-circuit-domain, and digital-domain approaches. In such a system, the high-performance channel estimator becomes particularly important in order to dramatically reduce the effect of SI\cite{Dongkyu,Rongyi,Day}. Channel estimation and pilot design have been widely and deeply investigated in conventional TDD/FDD mode\cite{Edfors,Edfors2,Coleri,Shufe,Kang}. However, channel estimation and pilot designing in FD OFDM systems is a challenging problem and should be restudied due to the existence of full-duplex self-interference. A digitally assisted analog channel estimator is designed to estimate SI channel for in-band FD radios\cite{Liu}. The authors in \cite{Koohian} propose two blind channel estimators to do simultaneous estimation for the SI and intended channels in FD wireless systems based on the expectation maximization and minimum mean square error approach. Using the maximum-likelihood criterion, the SI and intended channels are jointly estimated with the known transmitted symbols from itself and the pilot symbols from intended transceiver\cite{Masmoudi}. An iterative procedure is constructed to further enhance the estimation performance in the high signal-to-noise ratio (SNR) region \cite{Ahmed}. By using an adaptive orthogonal matching pursuit scheme, an time-domain least squares (TD-LS) channel estimator is proposed by exploiting the sparsity of SI channel and intended channel and measuring their sparisties \cite{Yuhai}. Considering IQ imbalances, a frequency-domain and DFT-based least-squares (LS) channel estimators are presented and the corresponding optimal pilot matrix product is proved to be an identity matrix multiplied by a constant\cite{Shufeng}. Also, the sum-MSE performance gain of the DFT-based LS channel estimator over the frequency-domain LS one is derived to be $N/L$ in white Gaussian noise scenario, where $N$ is the total number of subcarriers and $L$ is the length of cyclic prefix (CP).

How about the sum-MSE performance gain in the colored interference/noise scenarios? Is still it equal to $N/L$?  In this paper, we probe deeply into the trend of the sum-MSE performance gain in the presence of colored noise/interference in more detail. In the first step, the self-interference and intended channels are jointly estimated by the frequency-domain LS (FD-LS) and DFT-based LS channel estimators. Secondly, we derive their MSE expressions of the two channel estimators.  Then, we define the sum-MSE performance gain, whose exact expression is given. Avoiding the use of its cumbersome expression, we derive its simple upper and lower bounds by using the matrix eigen-value inequalities. Finally, by numerical simulation and  theoretical analysis, we find: the achievable sum-MSE performance gain ranges from 1 to $N/L$, and the upper bound is tighter than the lower bound in several typical extreme scenarios. The former is a good approximation to the exact sum-MSE performance gain.

This paper is organized as follows. The full-duplex system model is described in Section II. In Section III,  the FD-LS and DFT-based LS channel estimators are adopted to estimate both intended and self-interference channels, and their MSEs are derived. In Section IV, the sum-MSE performance gain  of the DFT-based channel estimator over the FD-LS LS one is defined, and its upper bound and lower bounds are derived. Simulation results and discussions are presented in Section V. Finally, Section VI concludes this whole paper.

\emph{Notations:} throughout the paper, matrices and vectors are denoted by letters of bold upper case and bold lower case, respectively. Signs $(\bullet)^H$, $(\bullet)^*$, $(\bullet)^T$, $(\bullet)^{-1}$, $\text{tr}(\bullet)$, $\Vert\bullet\Vert_{F}$, and $\text{det}(\bullet)$ denote matrix conjugate transpose, conjugate, transpose, inverse, trace, norm-2, and determinant, respectively. The notation $\mathcal E\{\bullet\}$ refers to the expectation operation. The symbol $\mathbf{I}_{n}$  denotes the $n \times n$ identity matrix. $\mathbf{0}_{n\times m}$ denotes an all-zero matrix of size $n\times m$.
 %$\otimes$ denotes the Kronecker product of two matrices. $\text{vec}(\mathbf{X})$ is an operation of stacking all columns of  $\mathbf{X}$ to a large column vector.
 %$\text{diag}\left\{\mathbf{a}\right\}$ denotes  an operation of placing  all elements of the vector $\mathbf{a}$ over the diagonal of diagonal matrix.

%======================Section II System Model======================
\section{System Model}
\begin{figure}[h]
  \centering
  \includegraphics[scale=0.46]{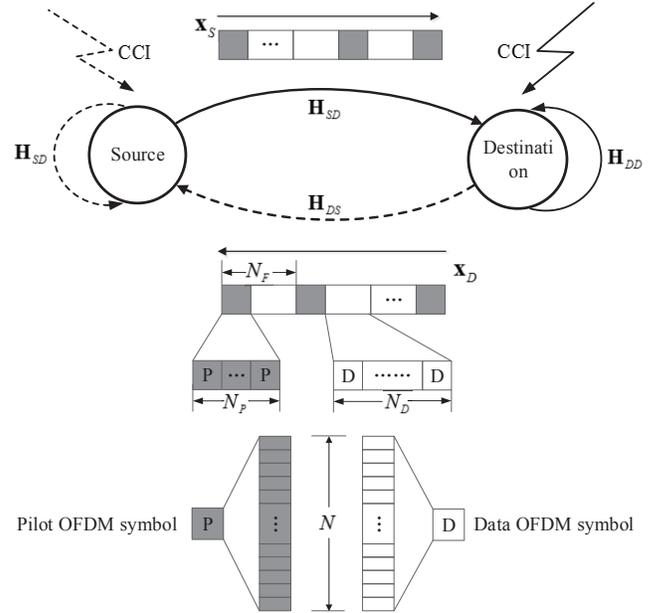}\\
  \caption{Block diagram of full-duplex OFDM system model.}
  \label{fig:1}
\end{figure}

 Fig.~\ref{fig:1} plots the block diagram of a point-to-point full-duplex OFDM system. Here, the destination node is used as a reference. The received vector at destination node is the summation of the signal from source node via intended channel $\mathbf{H}_{SD}$, the signal from local transmitter via SI channel $\mathbf{H}_{DD}$, and the co-channel interference (CCI) from other nodes. Both channels from source to destination (S2D) and destination to destination (D2D) are assumed to be time-invariant within one frame, where each frame consists of $N_F$ OFDM symbols, but vary from one frame to another. Each frame consists of $N_P$ pilot OFDM symbols and $N_D$ data OFDM symbols, which is shown in Fig.~\ref{fig:1}. Usually, in a practical system, $N_D$ is taken to be far larger than $N_P$ to achieve a high-spectrum efficiency. As shown in Fig.~1, block-type pilot pattern is adopted to estimate both D2D and S2D  unknown channels.

 Similar to \cite{ShuF}, the ideal channel frequency responses (CFR) has the following relationship with its channel impulse responses (CIR)
    \begin{align}
   \label{TD2FD_1}
    \mathbf{H}_{SD}&=\mathbf{F}_{N\times N}\left(\begin{array}{c}
                                         \mathbf{h}_{SD}\\
                                         \mathbf{0}_{(N-L)\times 1}
                                       \end{array}\right)
                                       =\mathbf{F}_{N\times L}\mathbf{h}_{SD},
    \end{align}
   \begin{align}
   \label{TD2FD_2}
    \mathbf{H}_{DD}&=\mathbf{F}_{N\times N}\left(\begin{array}{c}
                                         \mathbf{h}_{DD}\\
                                         \mathbf{0}_{(N-L)\times 1}
                                       \end{array}\right)
                                       =\mathbf{F}_{N\times L}\mathbf{h}_{DD},
    \end{align}
where $\mathbf{h}_{SD}$ and $\mathbf{h}_{DD}$ are  the  S2D and D2D CIRs defined by
\begin{align}
\mathbf{h}_{SD}=[h_{SD}(1)~h_{SD}(2)~ \cdots~ h_{SD}(L)]^T,
\end{align}
and
\begin{align}
\mathbf{h}_{DD}=[h_{DD}(1)~h_{DD}(2)~ \cdots~ h_{DD}(L)]^T,
\end{align}
respectively. $N$ is the total number of subcarriers, $L$ is the length of the CP, and $\mathbf{F}_{N\times N}$ is the normalized discrete Fourier transform matrix as
\begin{equation}
    \mathbf{F}_{N\times N}=\frac{1}{\sqrt{N}}\left(\begin{array}{cccc} 1 & 1 & \cdots & 1 \\

                                                                   1 & W^1 & \cdots & W^{N-1}\\
                                                                 \vdots & \vdots & \ddots & \vdots \\
                                                                 1 & W^{N-1} & \cdots & W^{(N-1)(N-1)}

     \end{array}\right)
\end{equation}
with $W=e^{-j\frac{2\pi}{N}}$.

The transmit vectors corresponding to the $n$th OFDM symbol from source and destination are denoted by
    \begin{align}
    \mathbf{x}_S(n,:)&=\left(x_S(n,1)~x_S(n,2)~\cdots~x_S(n,N)\right)^T,
    \end{align}
and
    \begin{align}
    \mathbf{x}_D(n,:)&=\left(x_D(n,1)~x_D(n,2)~\cdots~x_D(n,N)\right)^T,
    \end{align}
with average power $P_S$ and $P_D$ of each subcarrier, respectively.
After experiencing multipath channel $\mathbf{H}_{SD}$ and $\mathbf{H}_{DD}$, the received signal corresponding to the $n$th OFDM symbol at destination node is modeled as
\begin{align}
    \label{sys_mod_vec}
    \mathbf{y}(n,:)&=
        \text{diag}\{\mathbf{x}_S(n,:)\}\mathbf{H}_{SD}+\text{diag}\{\mathbf{x}_D(n,:)\}\mathbf{H}_{DD}\\ \nonumber
        &={\left(
            \begin{array}{cc}
              \text{diag}\{\mathbf{x}_S(n,:)\} & \text{diag}\{\mathbf{x}_D(n,:)\} \\
            \end{array}
          \right)
          \left(
            \begin{array}{c}
              \mathbf{H}_{SD}  \\
              \mathbf{H}_{DD}  \\
            \end{array}
          \right)}
        \\ \nonumber
        &+\mathbf{w}_{cci}(n,:)+\mathbf{w}_n(n,:),
    \end{align}
where $\mathbf{w}_{cci}(n,:)$ denotes the co-channel interference vector and $\mathbf{w}_n(n,:)$ is the noise vector in frequency domain. For convenience of the following derivation and analysis, the sum of co-channel interference and additive noise vector will be viewed as the new colored interference-plus-noise vector as follows
\begin{align}
\mathbf{w}(n,:)=\mathbf{w}_{cci}(n,:)+\mathbf{w}_n(n,:)
\end{align}
where the above interference-plus-noise vector is assumed to be independent along time direction $n$ \cite{Mishra}.
 %The temporal covariance matrix of $\mathbf{w}_i(1:N_P,k)$ over $k$th subcarrier is defined as
%\begin{align}
%\mathbf{R}_t=\mathcal E\{\mathbf{w}_i(1:N_P,k)\mathbf{w}_i(1:N_P,k)^H\}=\sigma_I^2\mathbf{I}_{N_P}
%\end{align}

Obviously, Eq.~(\ref{sys_mod_vec}) is an under-determined linear equation. Thus at least two pilot OFDM symbols are required for source node and destination node to estimate the unknowns containing both $\mathbf{H}_{SD}$ and $\mathbf{H}_{DD}$. Below, $N_P\geq 2$ consecutive pilot OFDM symbols are utilized to do one-time channel estimation:
\begin{align}
    \label{sys_mod_vect}
    &\left(\begin{array}{c}
      \mathbf{y}(n,:) \\
      \mathbf{y}(n+1,:) \\
      \vdots\\
      \mathbf{y}(n+N_P-1,:)
    \end{array}\right)\\ \nonumber
    &=\left(
            \begin{array}{cc}
            \text{diag}\{\mathbf{x}_S(n,:)\} & \text{diag}\{\mathbf{x}_D(n,:)\} \\
            \text{diag}\{\mathbf{x}_S(n+1,:)\} & \text{diag}\{\mathbf{x}_D(n+1,:)\} \\
            \vdots & \vdots \\
            \text{diag}\{\mathbf{x}_S(n+N_P-1,:)\} & \text{diag}\{\mathbf{x}_D(n+N_P-1,:)\}
            \end{array}
          \right)\\ \nonumber
     &\left(
            \begin{array}{c}
              \mathbf{H}_{SD}  \\
              \mathbf{H}_{DD}  \\
            \end{array}
          \right)
          +\underbrace{\left(\begin{array}{c}
                \mathbf{w}(n,:)\\
                \mathbf{w}(n+1,:)\\
                \vdots\\
                \mathbf{w}(n+N_P-1,:)\\
          \end{array}\right)}_{\tilde{\mathbf{w}}},
    \end{align}
    Considering the above interference-plus-noise $\tilde{\mathbf{w}}$ is colored and assuming its correlation function along time and frequency are independent, the covariance matrix of colored interference and noise is written  in the following form
\begin{align}
\mathbf{R}_{CCI}=\mathcal{E}\left\{\tilde{\mathbf{w}}\tilde{\mathbf{w}}^H\right\}=\mathbf{R}_t\otimes \mathbf{R}_w
\end{align}
where matrix $\mathbf{R}_w$ denotes the $N\times N$  frequency-domain  covariance matrix given by
\begin{align}
\mathbf{R}_w=\left(\begin{array}{cccc}
                     r_w(0)  & r_w(-1) & \cdots & r_w(1-N) \\
                     r_w(1) & r_w(0) & \cdots & r_w(2-N)\\
                     \vdots & \vdots & \ddots & \vdots \\
                     r_w(N-1) & r_w(N-2) & \cdots &  r_w(0)
 \end{array}\right),
 \end{align}
and  matrix $\mathbf{R}_t$ denotes the $N_P\times N_P$ time-direction covariance matrix given by
\begin{align}
\mathbf{R}_t=\left(\begin{array}{cccc}
                     r_t(0)  & r_t(-1) & \cdots & r_t(1-N_P) \\
                     r_t(1) & r_t(0) & \cdots & r_t(2-N_P)\\
                     \vdots & \vdots & \ddots & \vdots \\
                     r_t(N_P-1) & r_t(N_P-2) & \cdots &  r_t(0)
 \end{array}\right)
\end{align}
with $r_w(\Delta k)=\mathcal{E}\left\{w(n,k)w(n,k+\Delta k)^*\right\}$ and $r_t(\Delta n)=\mathcal{E}\left\{w(n,k)w(n+\Delta n,k)^*\right\}$.
%Regardless Of Frequency-Domain Correlation, Its Time-Direction Covariance Matrix Per Subchannel Is Defined As
%\begin{align}
%\mathbf{R}_{t}=\mathcal{E}\left\{\mathbf{w}(:,k)\mathbf{w}(:,k)^H\right\}
%\end{align}
%\begin{align}
%=\left(\begin{array}{cccc}
%                                                             r_t(1,1)  & r_t(1,2) & \cdots & r_t(1,N_P) \\
%                                                                   r_t(2,1) & r_t(2,2) & \cdots & r_t(2,N_P)\\
%                                                                 \vdots & \vdots & \ddots & \vdots \\
%                                                                 r_t(N_P,1) & r_t(N_P,2) & \cdots &  r_t(N_P,N_P)
% \end{array}\right)
%\end{align}
%where

\section{Frequency-domain LS and DFT-based LS channel estimators}

In this section,  we derive a closed-form expression for the optimal pilot matrix per subcarrier.
For the convenience of deriving below, we extract the $k$th subcarrier of the received symbol as follows
\begin{align}
\label{sys_mod_k}
    \mathbf{y}_{k}=\mathbf{P}_k\mathbf{H}_k+\mathbf{w}_k,
\end{align}
with
\begin{align}
    \mathbf{y}_{k}=\left(\begin{array}{ccc}
                    y(n,k) & \cdots &  y(n+N_P-1,k)
                   \end{array}\right)^T,
\end{align}

\begin{align}
    \mathbf{P}_k=\left(
                    \begin{array}{cc}
                    x_S(n,k) & x_D(n,k) \\
                    \vdots & \vdots \\
                    x_S(n+N_P-1,k) & x_D(n+N_P-1,k)
                    \end{array}\right),
\end{align}
\begin{align}\label{H k}
    \mathbf{H}_{k}=\left(\begin{array}{cc}
                    H_{SD}(k) &  H_{DD}(k)
                   \end{array}\right)^T,
\end{align}
and
\begin{align}
\mathbf{w}_k=\left(\begin{array}{ccc}
                w(n,k) & \cdots & w(n+N_P-1,k)
              \end{array}\right)^T.
\end{align}
Note that the $\mathbf{H}_{k}$ in (\ref{H k}) is just the channel parameter to be estimated.

In terms of (\ref{sys_mod_k}), FD-LS estimator can be given by
\begin{align}\label{FD_LS}
    \hat{\mathbf{H}}_k=\left(\mathbf{P}_k^H\mathbf{P}_k\right)^{-1}\mathbf{P}_k^H\mathbf{y}_{k}=\mathbf{H}_k+\left(\mathbf{P}_k^H\mathbf{P}_k\right)^{-1}\mathbf{P}_k^H\mathbf{w}_k
\end{align}
Let us define the channel estimation error as
\begin{align}
    \Delta\mathbf{H}_k=\left(\mathbf{P}_k^H\mathbf{P}_k\right)^{-1}\mathbf{P}_k^H\mathbf{w}_k
\end{align}
which forms the MSE of the FD-LS estimator over subcarrier $k$ as follows
\begin{align}
    \text{MSE}_k&=\mathcal{E}\left\{\text{tr}\left[\Delta\mathbf{H}_k(\Delta\mathbf{H}_k)^H\right]\right\}\\ \nonumber
       &=\text{tr}\left[\mathbf{P}_k\left(\mathbf{P}_k^H\mathbf{P}_k\right)^{-2}\mathbf{P}_k^H\mathcal{E}\left(\mathbf{w}_k\mathbf{w}_k^H\right)\right]
\end{align}

Since interference-plus-noise is assumed to be independent across different OFDM symbols, the temporal covariance matrix will be
\begin{align}
    \mathbf{R}_t=\mathcal{E}\left(\mathbf{w}_k\mathbf{w}_k^H\right)=\sigma_{I}^2\mathbf{I}_{N_P}
\end{align}
where $\sigma_{I}^2$ represents the average power of interference-plus-noise\cite{Mishra}. Thus, the MSE will be simplified as
\begin{align}
    \text{MSE}_k&=\sigma_{I}^2\text{tr}\left[\left(\mathbf{P}_k^H\mathbf{P}_k\right)^{-1}\right]
\end{align}
In order to optimize the performance of FD-LS channel estimator, we should minimize the above $\text{MSE}_k$ by designing the pilot matrix $\mathbf{P}_k$ with the constraint of transmit power of source and destination nodes, which can be expressed as the following optimization problem
\begin{align}
\text{min}~~~&~~\text{tr}\left[\left(\mathbf{P}_k^H\mathbf{P}_k\right)^{-1}\right]\\ \nonumber
\text{s.t.}~~&(\mathbf{P}_k^H\mathbf{P}_k)_{11}\leq N_PP_S\\ \nonumber
             &(\mathbf{P}_k^H\mathbf{P}_k)_{22}\leq N_PP_D
\end{align}
where $P_S$ and $P_D$ are the average transmit power per subcarrier of source and destination nodes, respectively.

Define $\mathbf{X}=\mathbf{P}_k^H\mathbf{P}_k$ and use the property of trace operator, the optimal optimization problem is relaxed into
\begin{align}\label{Opt_OneContraint}
\text{min}~~~&~~\text{tr}(\mathbf{X}^{-1})\\ \nonumber
\text{s.t.}~~&\text{tr}(\mathbf{X})\leq N_PP_S+N_PP_D.
\end{align}
To solve the above convex optimization problem, we define the associated Lagrangian function
\begin{align}
f(\mathbf{X},\lambda)=\text{tr}(\mathbf{X}^{-1})+\lambda(\text{tr}(\mathbf{X})-N_PP_S-N_PP_D).
\end{align}
Setting the first-order derivative of the above function with respect to $\mathbf{X}$ to zero, we have
\begin{align}\label{Tr_D_Der}
\frac{\partial f(\mathbf{X},\lambda)}{\partial\mathbf{X}}=\frac{\partial\text{tr}(\mathbf{X}^{-1})}{\partial\mathbf{X}}+\lambda\mathbf{I}_2=\mathbf{0}.
\end{align}
In accordance with the proof in Appendix A, we have
\begin{align}
\frac{\partial\text{tr}(\mathbf{X}^{-1})}{\partial\mathbf{X}}=-(\mathbf{X}^{-2})^T.
\end{align}
Inserting the above expression in the right-hand side of (\ref{Tr_D_Der}) gives
\begin{align}
\frac{\partial f(\mathbf{X},\lambda)}{\partial\mathbf{X}}=-(\mathbf{X}^{-2})^T+\lambda\mathbf{I}_2=\mathbf{0},
\end{align}
which means
\begin{align}
\mathbf{X}^{-2}=\lambda\mathbf{I}_2,
\end{align}
It can be further reduced towards
\begin{align}\label{Optimal_PilotPattern}
    \mathbf{X}=\frac{1}{\sqrt{\lambda}}\mathbf{I}_2,
\end{align}
which is called the optimal pilot condition.
Using the power constraint in (\ref{Opt_OneContraint}), we have
\begin{align}
    \sqrt{\lambda}=\frac{2}{N_P(P_S + P_D)},
\end{align}
then  the optimal condition (\ref{Optimal_PilotPattern}) is represented as
\begin{align}
\label{Optimal_Pilot_condition}
\mathbf{P}_k^H\mathbf{P}_k=\frac{N_P(P_S + P_D)}{2}\mathbf{I}_2
\end{align}
where $\mathbf{P}_k$ is chosen to be any two columns of $N_P\times N_P$ unitary matrix multiplied by any predefined constant. For example, given $N_P=4$ and 16QAM constellation, the optimal $\mathbf{P}_k$ is designed as follows
\begin{align}
{\mathbf{P}_k}^{\star}=\frac{\sqrt{(P_S+P_D)}(1+3i)}{2\sqrt{5}}\left(\begin{array}{cc}
                 1 & ~1 \\
                 1 & ~1 \\
                 1 & ~-1\\
                 1 & ~-1\\
                               \end{array}\right).
\end{align}

\section{Sum-MSE performance gain derivation and analysis}
In this section, we derive the sum-MSE performance expressions of both FD-LS and DFT-based LS channel estimators. The sum-MSE performance gain is defined as the ratio of the MSE of FD-LS channel estimator to that of DFT-based one. Its upper bound and lower bound are derived jointly. In two extreme situations: independent and full-correlated,  the simple expressions of the corresponding sum-MSE performance gains are directly given and discussed.

When the optimal pilot matrix satisfying (\ref{Optimal_Pilot_condition}) is adopted, the FD-LS estimator in (\ref{FD_LS}) will become
\begin{align}
\hat{\mathbf{H}}_k=\mathbf{H}_k+\frac{2}{N_P(P_S+P_D)}\mathbf{P}_k^{H}\mathbf{w}_k,
\end{align}
which is equivalently written in the following form
\begin{align}
\hat{H}_{SD}(k)=&H_{SD}(k)+\\ \nonumber
                &\frac{2}{N_P(P_S+P_D)}\sum_{p=0}^{N_P-1}x_S^*(n+p,k)w(n+p,k),
\end{align}
and
\begin{align}
\hat{H}_{DD}(k)=&H_{DD}(k)+\\ \nonumber
                &\frac{2}{N_P(P_S+P_D)}\sum_{p=0}^{N_P-1}x_D^*(n+p,k)w(n+p,k).
\end{align}
Due to the similar forms of the above two equations, we take $\mathbf{H}_{DD}$ as an example below. Stacking all the subcarriers, we can model the estimated channel gain vector as follows
\begin{align}\label{FD-LS-Err-Mod}
\hat{\mathbf{H}}_{DD}=\mathbf{H}_{DD}+\mathbf{e}_{DD},
\end{align}
where
\begin{align}
    \mathbf{e}_{DD}=\frac{2}{N_P(P_S+P_D)}\sum_{p=0}^{N_P-1}\text{diag}\{\mathbf{x}_D^*(n+p,:)\}\mathbf{w}(n+p,:)
\end{align}
denotes the estimation error due to the FD-LS estimator. Thus the corresponding MSE of $\mathbf{H}_{DD}$ is given by
\begin{align}
\text{MSE}_{DD}=\mathcal{E}\left(\text{tr}\left\{\mathbf{e}_{DD}\left(\mathbf{e}_{DD}\right)^H\right\}\right)=\text{tr}\left\{\mathcal{E}\left(\mathbf{e}_{DD}\left(\mathbf{e}_{DD}\right)^H\right)\right\}.
\end{align}
Due to
\begin{align}
\mathcal{E}\left(\mathbf{x}_D^*(n+p,:)\mathbf{x}_D^*(n+p,:)^T\right)=P_D\mathbf{I}_N,
\end{align}
we have
\begin{align}
\label{autocor}
\mathcal{E}\left(\mathbf{e}_{DD}\left(\mathbf{e}_{DD}\right)^H\right)=\frac{2P_D\sigma_{I}^2}{(P_S+P_D)^2}\mathbf{A}
\end{align}
where $\mathbf{A}$ denotes the normalized frequency-domain covariance matrix of $\mathbf{w}(n,:)$ with
\begin{align}
\mathbf{R}_w=\sigma_{I}^2\mathbf{A}.
\end{align}
Obviously, the above $\mathbf{A}$ is a positive semi-definite Hermitian matrix and its diagonal elements are $1$. Therefore, the above MSE of $\mathbf{H}_{DD}$ can be simplified as
\begin{align}
\label{mse_fdls}
\text{MSE}_{DD}=\frac{2NP_D\sigma_{I}^2}{(P_S+P_D)^2}.
\end{align}
In the same manner, we can obtain the $\text{MSE}_{SD}$ corresponding to $\mathbf{H}_{SD}$ as follows
\begin{align}
\label{mse_fdls}
\text{MSE}_{SD}=\frac{2NP_S\sigma_{I}^2}{(P_S+P_D)^2}.
\end{align}
In terms of the above two MSEs, we define the sum-MSE performance of the FD-LS channel estimator as follows
\begin{align}\label{SumMSE_FD-LS}
\text{SumMSE}_{FD-LS}=\text{MSE}_{DD}+\text{MSE}_{SD}=\frac{2N\sigma_{I}^2}{P_S+P_D}.
\end{align}
Using the transform relationship in (\ref{TD2FD_2}), the corresponding estimated time-domain CIR channel gain vectors will be given by
\begin{align}
\label{FD2TD}
\hat{\mathbf{h}}_{DD}=\mathbf{E}_{L\times N}\mathbf{F}_{N\times N}^{H}\hat{\mathbf{H}}_{DD},
\end{align}
where
\begin{align}
\mathbf{E}_{L\times N}=\left(\begin{array}{cc}
                                        \mathbf{I}_L & \mathbf{0}_{L\times(N-L)}
                                        \end{array}\right).
\end{align}
Performing the DFT operations to both sides of (\ref{FD2TD}) yields the following DFT-based channel estimator
\begin{align}
\label{FD2FD}
\tilde{\mathbf{H}}_{DD}=\mathbf{F}_{N\times L}\mathbf{E}_{L\times N}\mathbf{F}_{N\times N}^{H}\hat{\mathbf{H}}_{DD}.
\end{align}
%Due to the above FFT/IFFT operation process of exploiting the time-domain property of CIR, the sum MSE of the improved channel estimator will be reduced.
Combining the error model in (\ref{FD-LS-Err-Mod}), we have the estimation error model of the DFT-based LS estimator
\begin{align}\label{DFT_ErrMod}
\tilde{\mathbf{H}}_{DD}&=\mathbf{F}_{N\times L}\mathbf{E}_{L\times N}\mathbf{F}_{N\times N}^{H}\hat{\mathbf{H}}_{DD}\\ \nonumber
&=\mathbf{H}_{DD}+\mathbf{F}_{N\times L}\mathbf{E}_{L\times N}\mathbf{F}_{N\times N}^{H}\mathbf{e}_{DD}.
\end{align}
In terms of (\ref{DFT_ErrMod}), the corresponding MSE of $\mathbf{H}_{DD}$ by DFT-based LS will be expressed as
\begin{align}
\text{MSE}'_{DD}&=\mathcal{E}\Big(\text{tr}\Big\{\left(\mathbf{e}_{DD}\right)^H\mathbf{F}_{N\times N}\mathbf{E}_{L\times N}^H\mathbf{F}_{N\times L}^H\mathbf{F}_{N\times L}\nonumber\\
&~~~~~~~~\mathbf{E}_{L\times N}\mathbf{F}_{N\times N}^{H}\mathbf{e}_{DD}\Big\}\Big)
\end{align}
Since $\mathbf{F}_{N\times N}\mathbf{E}_{L\times N}^H=\mathbf{F}_{N\times L}$ and $\mathbf{F}_{N\times L}^H\mathbf{F}_{N\times L}=\mathbf{I}_L$, the above MSE reduces to
\begin{align}
\label{MSE_DD_DFT_1}
\text{MSE}'_{DD}&=\mathcal{E}\Big(\text{tr}\Big\{\left(\mathbf{e}^{n}_{DD}\right)^H\mathbf{F}_{N\times L}\mathbf{F}_{N\times L}^{H}\mathbf{e}_{DD}\Big\}\Big)\nonumber\\
&=\text{tr}\Big\{\mathcal{E}\left(\mathbf{e}_{DD}\left(\mathbf{e}_{DD}\right)^H\right)\mathbf{F}_{N\times L}\mathbf{F}_{N\times L}^{H}\Big\}
\end{align}
Substituting (\ref{autocor}) into the above equation
\begin{align}
\label{MSE_DD_DFT_2}
\text{MSE}'_{DD}&=\text{tr}\Big\{\frac{2P_D\sigma_{I}^2}{(P_S+P_D)^2}\mathbf{A}\mathbf{F}_{N\times L}\mathbf{F}_{N\times L}^{H}\Big\}\\ \nonumber
&=\frac{2P_D\sigma_{I}^2}{(P_S+P_D)^2}\text{tr}\left\{\mathbf{A}\mathbf{B}\right\}
\end{align}
%In practical wireless communication systems, often the received signal is impaired by dominant interference sources. For example, in cellular systems, the dominant interference source can be a co-channel or an adjacent channel interferer. Then the estimation error $\mathbf{e}^{n}_{DD}$ is no longer independent and identically distributed, $\mathcal{E}\left(\mathbf{e}^{n}_{DD}\left(\mathbf{e}^{n}_{DD}\right)^H\right)=\mathbf{A}$
%where
%\begin{align}
%\mathbf{A}_{ij}=\mathcal{E}\left(e^{n}_{DD}(i)\left(e^{n}_{DD}(j)\right)^H\right)=\rho[i-j]\sigma_{DD,e}^2
%\end{align}
%obviously, $\mathbf{A}$ is Hermitian.
with
\begin{align}\label{B_Normal_Expr}
\mathbf{B}=\mathbf{F}_{N\times L}\mathbf{F}_{N\times L}^H,
\end{align}
and $\mathbf{B}$ can be further decomposed as
\begin{align}\label{B_expr}
\mathbf{B}=\mathbf{F}_{N\times N}\left(
            \begin{array}{cc}
            \mathbf{I}_{L\times L} & \mathbf{0}_{(N-L)\times L}  \\
            \mathbf{0}_{L\times (N-L)} & \mathbf{0}_{(N-L)\times (N-L)}  \\
            \end{array}
            \right)
\mathbf{F}_{N\times N}^H
\end{align}
which implies the singular value of matrix $\mathbf{B}$ is  0 or 1.
In the same manner, we can obtain MSE corresponding to $\mathbf{H}_{SD}$ as follows
\begin{align}
\label{MSE_SD_DFT}
\text{MSE}'_{SD}&=\text{tr}\Big\{\frac{2P_S\sigma_{I}^2}{(P_S+P_D)^2}\mathbf{A}\mathbf{F}_{N\times L}\mathbf{F}_{N\times L}^{H}\Big\}\\ \nonumber
&=\frac{2P_S\sigma_{I}^2}{(P_S+P_D)^2}\text{tr}\left\{\mathbf{A}\mathbf{B}\right\}.
\end{align}
Adding (\ref{MSE_DD_DFT_2}) and (\ref{MSE_SD_DFT}) forms the sum-MSE performance of the DFT-based LS channel estimaror as follows
\begin{align}\label{SUM_MSE_DFT}
\text{SumMSE}_{DFT}=\text{MSE}'_{DD}+\text{MSE}'_{SD}=\frac{2\sigma_{I}^2}{P_S+P_D}\text{tr}\left\{\mathbf{A}\mathbf{B}\right\}.
\end{align}
To evaluate the sum-MSE performance gain achieved by the DFT-based channel estimator over the FD-LS one, let us define
\begin{align}\label{SumMSE_gain}
\gamma=\frac{\text{SumMSE}_{FD-LS}}{\text{SumMSE}_{DFT}}.
\end{align}
In dB, the sum-MSE performance gain achieved by the DFT-based channel estimator is $10\log_{10}\gamma$ dB. Using the inequality (\ref{mse_6}), the above performance gain is bounded by the following double side approximation
\begin{align}
\frac{\sum_{i=N-L+1}^N \lambda_i(\mathbf{A})}{N} \leq \gamma^{-1} \leq \frac{\sum_{i=1}^L \lambda_i(\mathbf{A})}{N}.
\end{align}
%\begin{align}
%\label{mse_5}
%\sum_{i=N-L+1}^N \sigma_i(\mathbf{A})\sigma_{DD,e}^2\leq\text{MSE}'^{n}_{DD}\leq \sum_{i=1}^L \sigma_i(\mathbf{A})\sigma_{DD,e}^2
%\end{align}
\begin{theorem}
Matrices $\mathbf{A}$ and $\mathbf{B}$ are $N\times N$ positive semi-definite, where $\lambda_1(\mathbf{A})$,$\cdots$, $\lambda_N(\mathbf{A})$ denote the eigenvalues of matrix $\mathbf{A}$, arranged in nondecreasing order. If matrix $\mathbf{B}$ has
the following form
\begin{align}\label{B_EVD}
\mathbf{B}=\mathbf{U}_B\left(
            \begin{array}{cc}
            \mathbf{I}_{L\times L} & \mathbf{0}_{L\times (N-L)}  \\
            \mathbf{0}_{(N-L)\times L} & \mathbf{0}_{(N-L)\times (N-L)}  \\
            \end{array}
            \right)
\mathbf{U}_B^H,
\end{align}
then we have the following inequality
\begin{align}
\label{mse_5}
\sum_{i=N-L+1}^N \lambda_i(\mathbf{A})\leq\text{tr}\left(\mathbf{A}\mathbf{B}\right)=\text{tr}\left(\mathbf{B}\mathbf{A}\right)\leq \sum_{i=1}^L \lambda_i(\mathbf{A})
\end{align}
%due to the non-negative property of all eigen-values of matrix $\mathbf{A}$.
\emph{Proof:} Please see Appendix \ref{Append_A}.\hfill$\blacksquare$\\
\end{theorem}

As a result,  the sum-MSE of the DFT-based LS will be bounded by
\begin{align}
\label{mse_6}
\frac{2\sigma_{I}^2}{P_S+P_D}\sum_{i=N-L+1}^N \lambda_i(\mathbf{A})&\leq \text{SumMSE}_{DFT} \\ \nonumber
&\leq\frac{2\sigma_{I}^2}{P_S+P_D}\sum_{i=1}^L \lambda_i(\mathbf{A}).
\end{align}

In a practical wireless communication system,  the correlation factor of these interference and noise can be estimated. We will discuss how the different matrix $\mathbf{A}$ affects the MSE performance  gains achieved by the  DFT-based LS channel estimator in two extreme scenarios:

\textbf{Scenario 1}: when  $\mathbf{A}=\mathbf{I}_N$ , that is, $A_{ii}=1$ and $A_{ij}=0, \forall i\neq j$, we have
\begin{align}
\text{tr}\left\{\mathbf{A}\mathbf{B}\right\}=\text{tr}\left\{\mathbf{B}\right\} \nonumber
\end{align}
\begin{align}
=\text{tr}\left\{\left(
            \begin{array}{cc}
            \mathbf{I}_{L\times L} & \mathbf{0}_{(N-L)\times L}  \\
            \mathbf{0}_{L\times (N-L)} & \mathbf{0}_{(N-L)\times (N-L)}  \\
            \end{array}
            \right)
\mathbf{F}_{N\times N}^H\mathbf{F}_{N\times N}\right\} \nonumber
\end{align}
\begin{align}
=\text{tr}\left\{\left(
            \begin{array}{cc}
            \mathbf{I}_{L\times L} & \mathbf{0}_{(N-L)\times L}  \\
            \mathbf{0}_{L\times (N-L)} & \mathbf{0}_{(N-L)\times (N-L)}  \\
            \end{array}
            \right)\right\}=L,
\end{align}
then
\begin{align}
\text{SumMSE}_{DFT}=\frac{2L\sigma_{I}^2}{P_S+P_D}.
\end{align}
By utilizing the definition of  (\ref{SumMSE_gain}), we have
\begin{align}
\gamma=\frac{N}{L},
\end{align}
which means the sum-MSE of DFT-based LS channel estimator will be reduced to ${L}/{N}$ of that of the FD-LS one when the interference-plus-noise vector is independent identically distributed (i.i.d).

\textbf{Scenario 2}: Considering $\mathbf{A}$ is a matrix of all-ones
\begin{align}\label{Mat-all-one}
\mathbf{A}=\mathbf{1}_N\mathbf{1}_N^H,
\end{align}
where $\mathbf{1}_N$ is an  $N$-D column vector of all-ones. Such type of colored interference and noise can be expressed as
\begin{align}\label{Full_Corr}
\mathbf{w}=g\mathbf{1}_N
\end{align}
where $g$ is any random variable obeying some typical random distribution. Substituting matrix in (\ref{Mat-all-one}) in the trace $\text{tr}\left\{\mathbf{A}\mathbf{B}\right\}$, we have
\begin{align}
\label{eq_mse_col2}
\text{tr}\left\{\mathbf{A}\mathbf{B}\right\}=\text{tr}\left\{\mathbf{1}\mathbf{1}^H\mathbf{F}_{N\times L}\mathbf{F}_{N\times L}^H\right\}=\text{tr}\left\{\mathbf{t}\mathbf{t}^H\right\}
\end{align}
where $\mathbf{t}=\mathbf{1}^H\mathbf{F}_{N\times L}$ is given by
\begin{align}
t_{l}=\frac{1}{\sqrt{N}}\sum_{n=1}^N W^{(l-1)(n-1)}=\left\{
  \begin{array}{ll}
    \sqrt{N}, & \hbox{$l=1$;} \\
    0, & \hbox{$2\leq l\leq L$;}
  \end{array}
\right.
\end{align}
Substituting the above equation into (\ref{eq_mse_col2}) yields
\begin{align}
\text{tr}\left\{\mathbf{A}\mathbf{B}\right\}=N,
\end{align}
then
\begin{align}
\text{SumMSE}_{DFT}=\frac{2N\sigma_{I}^2}{P_S+P_D}.
\end{align}
Plugging the above expression and (\ref{mse_fdls}) into (\ref{SumMSE_gain}) gives
\begin{align}
\gamma=1,
\end{align}
which implies there is no sum-MSE performance gain achievable by the DFT-based LS channel estimator. In other words, the DFT-based LS channel estimator has the same sum-MSE performance as the FD-LS one.
From the above two special scenarios, we conclude  that  the sum-MSE performance gains achieved by the DFT-based LS channel estimator are $10\log_{10}\frac{N}{L}$dB, and $0$dB in the independent and full-correlated cases, respectively. The correlation degree of covariance matrix of the colored interference and noise vector will impose a significant influence on the sum-MSE performance gain achieved by the DFT-based LS channel estimator. Obviously, the DFT-based LS channel estimator may  harvest a larger sum-MSE performance gain over the FD-LS one by increasing the value of $N/L$ in the case of i.i.d interference and noise.

%-------------------------------------Simulations and Discussions--------------------------------------------
\section{Simulations and Discussions}
In this section, we provide numerical results and analysis to evaluate the exact sum-MSE performance gain, its upper bound and its lower bound by changing the values of correlation factor $\rho$ and ratio $N/L$. %System parameters are set as follows: signal bandwidth $BW=10MHz$, digital modulation 16QAM, and carrier frequency $f_c=2GHz$. The typical EVA channel model with maximum path delay $2.51\mu$s in LTE standard is employed in our simulation.
%OFDM symbol length $N=512$, cyclic prefix $L=32$,

For simplicity, we choose an exponential correlation model to describe the frequency-domain covariance matrix $\mathbf{A}$ of interference-plus-noise as
\begin{align}
\mathbf{A}=\left(
             \begin{array}{ccccc}
               1 & \rho & \cdots & \rho^{N-1} \\
               \rho & 1 & \cdots & \rho^{N-2} \\
               \vdots & \vdots & \ddots & \vdots \\
               \rho^{N-1} & \rho^{N-2} & \cdots & 1 \\
             \end{array}
           \right)
\end{align}
where $\rho \in [0,1]$ denotes the frequency-domain correlation factor of colored interference-plus-noise vector. The correlation factor corresponding to two distinct subcarriers is defined as
\begin{align}
\rho^k=r_w(\Delta k)
\end{align}
where $\rho=0$ means that each element of the interference-plus-noise vector is independent identically distributed, and $\rho=1$ means the interference-plus-noise vector is full correlated. In other words, all elements of this vector obeys the same random distribution as shown in (\ref{Full_Corr}).
 \begin{figure}[h]
  \centering
  \includegraphics[scale=0.52]{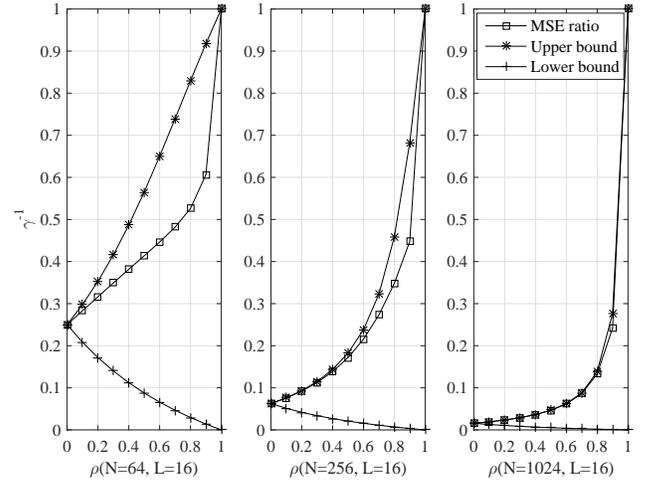}\\
  \caption{Curves of MSE ratio and its bounds versus $\rho$ with fixed $L=16$.}
  \label{fig:2}
\end{figure}

 \begin{figure}[h]
  \centering
  \includegraphics[scale=0.52]{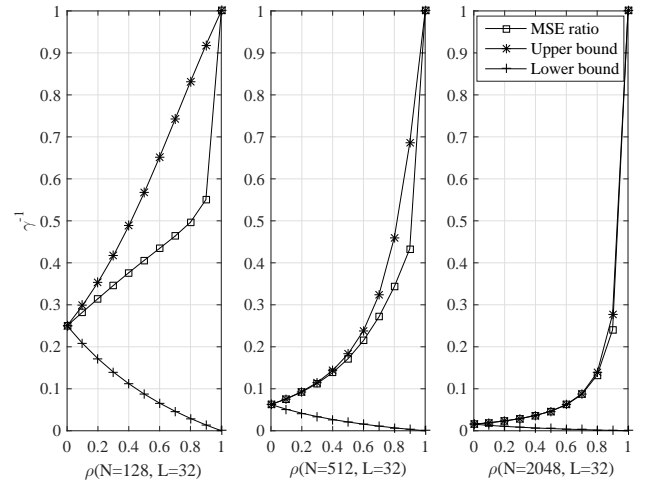}\\
  \caption{Curves of MSE ratio and its bounds versus  $\rho$ with fixed $L=32$.}
  \label{fig:3}
\end{figure}
Fig.~2 demonstrates the curves of $\gamma^{-1}$ and its bounds, including upper and lower bounds, versus $\rho$ for different ratios $N/L$ with fixed $L=16$. Observing this figure, we obtain the following results: given a fixed $L$, as $N$ increases, the upper bound converges to the exact value $\gamma^{-1}$. When both $L$ and $N$ are fixed, the lower bound is far away from the exact value $\gamma^{-1}$ with increase in the value of $\rho$. Consequently, we can make a conclusion that the upper bound is tighter compared to the lower bound.

Now, let us consider two extreme situations. At $\rho=1$, the upper bound and the exact value $\gamma^{-1}$ are equal to one. This means that there is no performance gain in the full-correlated-interference-plus-noise scenario. Conversely, at $\rho=0$, i.e., the case of i.i.d interference and noise vector, the MSE gain $\gamma$ equals $N/L$. This is also the achievable maximum performance gain by the DFT-based LS channel estimator. In particular, from Fig.~2, we also find that decreasing the value of $\rho$ will make an enhancement in the performance gain. Additionally, at two extreme-value points $(0,~L/N)$ and $(1,~1)$, the derived upper bound is in agreement with the exact sum-MSE value for all three subfigures of Fig.~2.

Fig.~3 illustrates the curves of $\gamma^{-1}$ and its bounds versus $\rho$ for different ratios $N/L$ with fixed $L=32$. It is evident that Fig.~3 shows the same performance as Fig.~2.

\begin{figure}[h]
  \centering
  \includegraphics[scale=0.52]{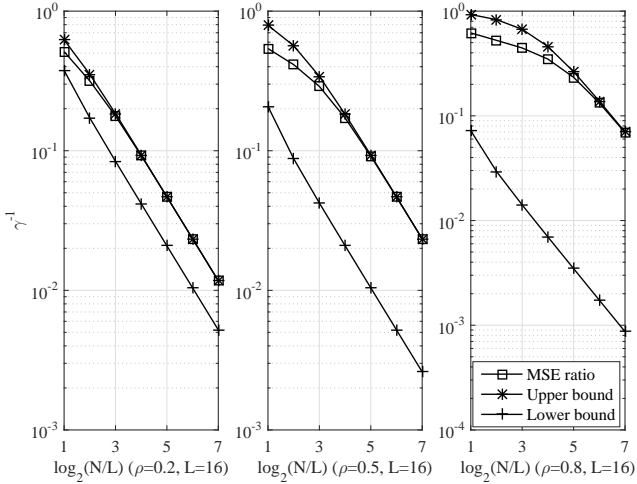}\\
  \caption{Curves of MSE ratio and its bounds versus  $\log_2(N/L)$ with fixed $L=16$.}
  \label{fig:4}
\end{figure}

 \begin{figure}[h]
  \centering
  \includegraphics[scale=0.52]{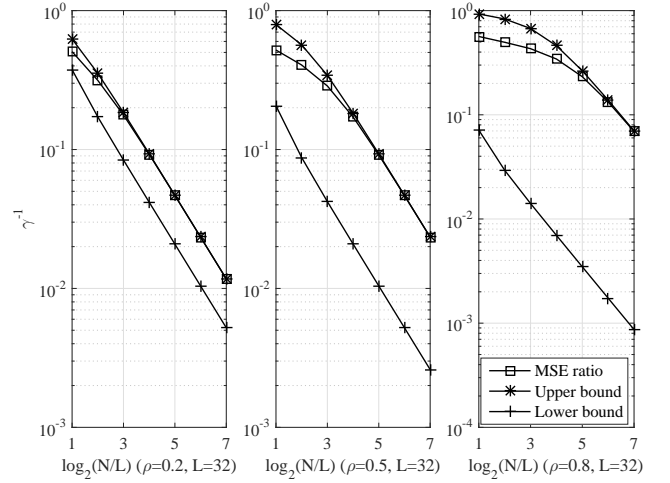}\\
  \caption{Curves of MSE ratio and its bounds versus  $\log_2(N/L)$ with fixed $L=32$.}
  \label{fig:5}
\end{figure}

Fig.~4 plots curves of $\gamma^{-1}$ and its bounds versus different ratios $\log_2(N/L)$ for different $\rho$ and fixed $L=16$. It intuitively follows from this figure that given $L$ and $\rho$,  as $N$ increases, $\gamma^{-1}$ will gets closer to its upper bound and even begins to overlap for a sufficiently large $N/L$. Specifically, for a smaller $\rho$, the curves of $\gamma^{-1}$ and its upper bound overlap with each other starting from the a relatively smaller $N$. For example, we can find that the curves of $\gamma^{-1}$ and its upper bound begin to overlap at the value of $N/L$ being 8, 16 and 64 for $\rho$ increases from 0.2, 0.5 to 0.8, respectively. Consequently, we may claim that the upper bound  shows a better approximation to the exact value of $\gamma^{-1}$ for almost all situations compared with lower bound.

Fig.~5 shows the curves of $\gamma^{-1}$ and its bounds versus different ratios $\log_2(N/L)$ for different values of correlation factor $\rho$ with fixed $L=32$, which yields the same trend as Fig.~4.

In summary, the values of $N/L$ and $\rho$  have a dramatic impact on the sum-MSE performance gain. Increasing the value of $N/L$  improves the sum-MSE performance gain whereas reducing the value brings the sum-MSE performance gain down.

%================================
\section{Conclusions}
In this paper, we investigate and analyze the sum-MSE performance gain of DFT-based LS channel estimator over FD-LS one in full-duplex OFDM system with colored interference and noise. The exact value, upper bound, and lower bound of the sum-MSE performance gain are derived, discussed, and verified. From numerical simulations and analysis, it follows that the upper bound is closer to the exact sum-MSE performance gain compared to the lower bound. In two extreme scenarios: full-correlated and independent interference plus noise vectors, the sum-MSE performance gain is shown to be $1$ and $N/L$, respectively. The correlation factor $\rho$ of interference plus noise vector  has a great impact on the sum-MSE performance gain. Roughly speaking, a small correlation factor $\rho$ will result in a large sum-MSE performance gain. Conversely, a large correlation factor $\rho$ will produce a small sum-MSE performance gain. The above results can be applied to provide a guidance for the design of channel estimation in future full-duplex wireless networks such as mobile communications, satellite communications, cooperative communications, V2V, unmanned-aerial-vehicles networks, and internet of things (IoT), etc.

\appendices
\section{Proof of Derivative of $\text{tr}(\mathbf{X}^{-1})$}
Using the complex differential property of trace operator \cite{Moon}, we have
\begin{align}\label{trace_op}
d\text{tr}(\mathbf{X}^{-1})=\text{tr}(d\mathbf{X}^{-1}).
\end{align}
Applying the differential operator to both sides of the identity
\begin{align}
\mathbf{X}^{-1}\mathbf{X}=\mathbf{I}
\end{align}
 yields
\begin{align}
(d\mathbf{X}^{-1})\mathbf{X}+\mathbf{X}^{-1}(d\mathbf{X})=\mathbf{0}.
\end{align}
Removing the second term of the left-hand side of the above equation to the right-hand side and rearranging forms
\begin{align}
d\mathbf{X}^{-1}=-\mathbf{X}^{-1}(d\mathbf{X})\mathbf{X}^{-1}.
\end{align}
Substituting the above equation into (\ref{trace_op}) gives
\begin{align}
d\text{tr}(\mathbf{X}^{-1})&=\text{tr}(-\mathbf{X}^{-1}(d\mathbf{X})\mathbf{X}^{-1})=-\text{tr}(\mathbf{X}^{-2}(d\mathbf{X}))\\ \nonumber
&=-\sum_{i}\sum_{j}(\mathbf{X}^{-2})_{ji}dX_{ij}.
\end{align}
The derivative of $\text{tr}(\mathbf{X}^{-1})$ with respect to element $X_{ij}$ is
\begin{align}
\frac{\partial\text{tr}(\mathbf{X}^{-1})}{\partial X_{ij}}=\frac{d\text{tr}(\mathbf{X}^{-1})}{dX_{ij}}=-(\mathbf{X}^{-2})_{ji}.
\end{align}
Therefore, the partial derivative of $\text{tr}(\mathbf{X}^{-1})$ with respect to $\mathbf{X}$ can be expressed as
\begin{align}
\frac{\partial\text{tr}(\mathbf{X}^{-1})}{\partial \mathbf{X}}=-(\mathbf{X}^{-2})^T.
\end{align}
\hfill$\blacksquare$

\section{Proof of Theorem 1}
Substituting (\ref{B_EVD}) in $\text{tr}\left(\mathbf{A}\mathbf{B}\right)$ yields

\begin{align}\label{A_B_EVD}
\text{tr}\left(\mathbf{A}\mathbf{B}\right)=\text{tr}\left(\mathbf{U}_B^H\mathbf{A}\mathbf{U}_B\Lambda_B\right)
\end{align}
where
\begin{align}\label{Lambda_B}
\mathbf{\Lambda_B}=\left(
            \begin{array}{cc}
            \mathbf{I}_{L\times L} & \mathbf{0}_{L\times (N-L)}  \\
            \mathbf{0}_{(N-L)\times L} & \mathbf{0}_{(N-L)\times (N-L)}  \\
            \end{array}
            \right)
,
\end{align}
Let us define
\begin{align}
\mathbf{\tilde{A}}=\mathbf{U}_B^H\mathbf{A}\mathbf{U}_B,
\end{align}
where $\mathbf{\tilde{A}}$ has the same set of eigen-values as  $\mathbf{A}$ due to the property of unitary transformation. Furthermore,
\begin{align}\label{tA_Lamdba_B}
\text{tr}\left(\mathbf{A}\right)=\text{tr}\left(\mathbf{\tilde{A}}\right)
\end{align}

The identity (\ref{A_B_EVD}) is rewritten as
\begin{align}\label{Tr_A_Lamdba_B}
\text{tr}\left(\mathbf{A}\mathbf{B}\right)=\text{tr}\left(\mathbf{\tilde{A}}\Lambda_B\right)
\end{align}
Similar to (\ref{Lambda_B}), $\mathbf{\tilde{A}}$ is represented in the block matrix
\begin{align}
\mathbf{\tilde{A}}=\left(
            \begin{array}{cc}
            \mathbf{\tilde{A}}_{11} &   \mathbf{\tilde{A}}_{21} \\
              \mathbf{\tilde{A}}_{12} &   \mathbf{\tilde{A}}_{22}  \\
            \end{array}
            \right),
\end{align}
Using the above expression,
\begin{align}\label{tA_Lamdba_B}
\text{tr}\left(\mathbf{A}\mathbf{B}\right)=\text{tr}\left(\mathbf{\tilde{A}}_{11}\right)
\end{align}
Making use of Theorem 4.3.28 in \cite{Horn}, we have
\begin{align}\label{tA_Lamdba_B}
\begin{array}{c}
  \lambda_{N-L+1}\left(\mathbf{\tilde{A}}\right) \leq \lambda_1\left(\mathbf{\tilde{A}}_{11}\right) \leq \lambda_1\left(\mathbf{\tilde{A}}\right) \\
  \lambda_{N-L+2}\left(\mathbf{\tilde{A}}\right) \leq \lambda_2\left(\mathbf{\tilde{A}}_{11}\right) \leq \lambda_2\left(\mathbf{\tilde{A}}\right) \\
  \vdots \\
  \lambda_N\left(\mathbf{\tilde{A}}\right) \leq \lambda_L\left(\mathbf{\tilde{A}}_{11}\right)\leq \lambda_L\left(\mathbf{\tilde{A}}\right)\\
 \\
\end{array}
\end{align}
Adding all the above $L$ inequalities results in the fact that the trace in (\ref{Tr_A_Lamdba_B}) may be bounded by
\begin{align}\label{Tr_Lower_Bound}
\sum_{i=N-L+1}^{N}\lambda_i\left(\mathbf{A}\right)\leq\sum_{i=1}^{L}\lambda_i\left(\mathbf{\tilde{A}}_{11}\right)=\text{tr}\left(\mathbf{\tilde{A}}_{11}\right)\leq \sum_{i=1}^{L}\lambda_i\left(\mathbf{A}\right).
\end{align}
%Making use of Theorem 7.4.1.1 (von Neumann) in \cite{Horn}, the trace in (\ref{Tr_A_Lamdba_B}) may be upper bounded by
%\begin{align}\label{Tr_Upper_Bound}
%\text{tr}\left(\mathbf{A}\mathbf{B}\right)=\text{tr}\left(\mathbf{\tilde{A}}\Lambda_B\right)\leq \sum_{l=N-L+1}^{N}\lambda_l\left(\mathbf{A}\right).
%\end{align}

\hfill$\blacksquare$\label{Append_A}

\bibliographystyle{IEEEtran}
\bibliography{IEEEabrv,FDIQ}
\end{document}